\documentclass[conference]{IEEEtran}
\IEEEoverridecommandlockouts
\usepackage{cite}
\usepackage{amsmath,amssymb,color,graphicx,caption,subcaption,comment,tikz,url,latexsym} 
\usepackage{algorithmic}
\usepackage{graphicx}
\usepackage{textcomp}
\usepackage{pgfplots}
\usepackage{xcolor}
\usepackage{url}
\usepackage{multirow}
\usepackage{mathtools}

\usepgfplotslibrary{units}
\usetikzlibrary{spy,backgrounds}
\usepackage{pgfplotstable}

\def\BibTeX{{\rm B\kern-.05em{\sc i\kern-.025em b}\kern-.08em
		T\kern-.1667em\lower.7ex\hbox{E}\kern-.125emX}}

\newtheorem{thm}{Theorem}
\newtheorem{example}{Example}
\newtheorem{definition}{Definition}
\newtheorem{lemma}{Lemma}

\newtheorem{corollary}{Corollary}

\usepackage{bbm}
\newcommand{\M}{\mathsf{M}}
\newcommand{\m}{\mathsf{m}}
\newcommand{\len}{\textrm{len}}

\newcommand{\Tx}{{T$_{X}$}}
\newcommand{\Rel}{{R$_{Y}$}}
\newcommand{\Rec}{{R$_{Z}$}}
\newcommand{\E}{\mathbb{E}}

\begin{document}
	\interdisplaylinepenalty=0
	\title{Optimal Exponents  In Cascaded Hypothesis Testing under Expected Rate Constraints\\
	}
	\author{\IEEEauthorblockN{Mustapha Hamad}
		\IEEEauthorblockA{\textit{LTCI, Telecom Paris, IP Paris} \\
			91120 Palaiseau, France\\
			mustapha.hamad@telecom-paris.fr}
		\and
		\IEEEauthorblockN{Mich\`ele Wigger}
		\IEEEauthorblockA{\textit{LTCI, Telecom Paris, IP Paris} \\
			91120 Palaiseau, France\\
			michele.wigger@telecom-paris.fr}
		\and
		\IEEEauthorblockN{Mireille Sarkiss}
		\IEEEauthorblockA{\textit{SAMOVAR, Telecom SudParis, IP Paris} \\
			91011 Evry, France\\
			mireille.sarkiss@telecom-sudparis.eu}
	}
	\allowdisplaybreaks[4]
	\sloppy
	\maketitle
	\begin{abstract}
Cascaded binary hypothesis testing is studied in this paper with two decision centers at the relay and the receiver. All terminals have their own observations, where we assume that the observations at the transmitter, the relay, and the receiver form a Markov chain in this order. The communication occurs over two hops, from the transmitter to the relay and from the relay to the receiver. Expected rate constraints are imposed on both communication links. In this work, we characterize the optimal type-II error exponents at the two decision centers under constraints on the allowed type-I error probabilities.  Our recent work characterized the optimal type-II error exponents in the special case when the two decision centers have same type-I error constraints and provided an achievability scheme for the  general setup. To obtain the exact characterization for the general case, in  this paper we provide a new  converse proof as well as  a new matching  achievability scheme. Our results indicate that under unequal type-I error constraints at the relay and the receiver, a  tradeoff  arises between the maximum type-II error probabilities at these two terminals. Previous results showed that such  a tradeoff does not exist under equal type-I error constraints or under general type-I error constraints when a maximum rate constraint is imposed on the communication links.

\end{abstract}
\begin{IEEEkeywords}
	Multi-hop, distributed hypothesis testing, error exponents, expected rate constraints, variable-length coding, 
\end{IEEEkeywords}	
\section{Introduction}

In a very connected world, where Internet of things (IoT) and sensor networks are emerging widely, distributed hypothesis testing has been utilized for improving decisions under communication constraints. A well-known application is the cascaded hypothesis testing where sensors communicate in a serial way forming a multi-hop network. We consider binary hypothesis testing over a two-hop network composed of a sensor, a relay, a receiver and two decision centers placed at the relay and the receiver. In such a setup, both decision centers try to correctly guess the binary hypothesis $\mathcal{H}\in\{0,1\}$ underlying all terminals' observations including their own. Each decision center aims to maximize the accuracy of its decisions, where the error under the alternative hypothesis $\mathcal{H}=1$ (called type-II error) is more critical than the error under the null hypothesis $\mathcal{H}=0$ (called type-I error). Specifically, both decision centers aim at maximizing the exponential decay (in the number of observed samples) of the type-II error probabilities under constraints on the accepted type-I error probabilities. 

While most information-theoretic works on distributed hypothesis testing constrain the \emph{maximum communication rates} between the terminals \cite{Ahlswede,Han,Amari,Wagner,Michele2,zhao2018distributed}, some recent works \cite{MicheleVLconf,JSAIT,HWS_ITW20,HWS21} have considered  \emph{expected rate constraints}. Expected rate constraints were first considered in \cite{MicheleVLconf, JSAIT} in a single-sensor single-decision center setup, and   the maximum error exponents were exactly characterized for  \emph{testing-against independence}  when under the alternative hypothesis, the observations are distributed according to the product of the distributions under the null hypothesis.   The optimal error exponent for this setup\cite{MicheleVLconf,JSAIT}, is achieved by a simple coding and decision scheme which chooses an  event  $\mathcal{S}_n$ of probability close  to the permissible type-I error probability $\epsilon$. Under this event, the transmitter   sends  a single bit to the decision center, allowing it to decide directly on the hypothesis $\mathcal{H}=1$. Otherwise, the transmitter and receiver run the optimal scheme under the maximum rate constraints \cite{Ahlswede,Han}. The described scheme achieves same type-II error exponent  as in \cite{Ahlswede,Han}, but with a communication rate that is reduced by the factor $(1-\epsilon)$. 
This gain is achieved by means of \emph{variable-length coding} which allows to send a message of different rate for each sequence observed at the transmitter. Notice that only under an expected rate constraint variable-length coding can improve performance, but not under maximum rate constraints. Similar conclusions also hold for more complicated setups, as we showed   in \cite{HWS_ITW20} for the partially-cooperating multi-access network with two sensors and a single decision center, and in \cite{HWS21} for a special case of the two-hop network studied in this paper.
\begin{figure}[htbp]
\vspace{-0.3cm}
	\centerline{\includegraphics[width=8cm, scale=0.65]{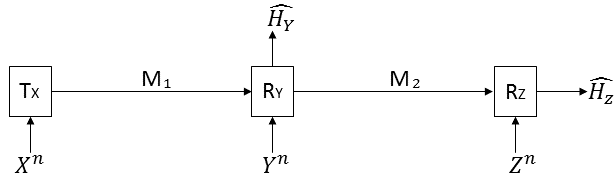}}\vspace{-0.05cm}
	\caption{Cascaded two-hop setup with two decision centers.}
	\label{fig:Cascaded}
\end{figure}

We consider the distributed hypothesis testing over the two-hop network in Figure~\ref{fig:Cascaded}, which consists of a transmitter, a relay, and a receiver, and where the observations at the transmitter $X^n$, the relay $Y^n$, and the receiver $Z^n$ form a Markov chain $X^n\to Y^n \to Z^n$ under both hypothesis.  Under maximum rate-constraints, the optimal type-II error exponents at the relay and the receiver for testing against independence  were characterized in \cite{Michele,Vincent}. Under expected rate constraints, \cite{HWS21} characterized the optimal type-II  error exponents  only when the relay and the receiver have same     type-I error constraint $\epsilon>0$. In this case, the maximum type-II error exponents can simultaneously be achieved at both terminals. Moreover, the expected rate constraints allow to boost both rates  by a factor $(1-\epsilon)^{-1}$ as compared to  maximum rate-constraints.
As in the single-user setup, the optimal exponents are achieved by a simple scheme where the transmitter chooses an event of probability $\epsilon$, and under this event both the transmitter and the relay send a single bit indicating the event to   the relay and the receiver, which then  decide on $\mathcal{H}=1$, and otherwise the optimal scheme of \cite{Michele} is run. 
For the general case, our previous work \cite{HWS21} only provides a set of achievable error exponents but no matching converse. 

In this paper, we provide an exact characterization of the optimal error exponents in the general case. We thus recover the main results of \cite{HWS21} as a special case. To obtain our results we   present both a new achievability result as well as a new converse proof. 

\textit{Notation:}
We follow the notation in \cite{ElGamal},\cite{JSAIT}. In particular, we use sans serif font for bit-strings: e.g., $\m$ for a deterministic and $\M$ for a random bit-string. We let  $\mathrm{string}(m)$ denote the shortest bit-string representation of a positive integer  $m$, and for any bit-string $\m$ we let  $\mathrm{len}(\m)$  and $\mathrm{dec}(\m)$ denote its length and its corresponding positive integer. In addition, $\mathcal{T}_{\mu}^{(n)}$ denotes the strongly typical set as defined in \cite[Definition 2.8]{Csiszarbook}.

	\section{System Model}

Consider the distributed hypothesis testing problem in Fig.~\ref{fig:Cascaded} under the Markov chain  
\begin{equation}\label{eq:Mc}
	X^n \to Y^n \to Z^n
\end{equation}
and  in the special case of testing against independence, i.e., depending on the binary hypothesis $\mathcal{H}\in\{0,1\}$, the tuple $(X^n,Y^n,Z^n)$ is distributed as:
\begin{subequations}\label{eq:dist}
	\begin{IEEEeqnarray}{rCl}
		& &\textnormal{under } \mathcal{H} = 0: (X^n,Y^n,Z^n) \sim \textnormal{i.i.d.} \, P_{XY}\cdot P_{Z|Y} ; \label{eq:H0_dist}\IEEEeqnarraynumspace\\
		& &\textnormal{under } \mathcal{H} = 1: (X^n,Y^n,Z^n) \sim \textnormal{i.i.d.} \, P_{X}\cdot P_{Y}\cdot P_{Z}
	\end{IEEEeqnarray} 
\end{subequations}
for given probability mass functions (pmfs) $P_{XY}$ and $P_{Z|Y}$.

The system consists of a transmitter T$_{X}$, a relay R$_{Y}$, and a receiver R$_Z$. The transmitter T$_{X}$ observes the source sequence $X^n$ and sends its bit-string message $\M_1 = \phi_1^{(n)}(X^n)$ to R$_Y$, where the encoding function is of the form $\phi_1^{(n)} : \mathcal{X}^n \to \{0,1\}^{\star}$ and satisfies the \emph{expected} rate constraint
\begin{equation}\label{eq:Rate1}
	\mathbb{E}\left[\mathrm{len}\left(\M_1\right)\right]\leq nR_1.
\end{equation} 
The relay R$_Y$ observes the source sequence $Y^n$ and with the message $\M_1$ received from T$_{X}$, it produces a guess  $\hat{\mathcal{H}}_Y$ of the hypothesis ${\mathcal{H}}$ using a decision function $g_1^{(n)} : \mathcal{Y}^n \times \{0,1\}^{\star} \to \{0,1\}$:
\begin{equation}
	\hat{\mathcal{H}}_Y = g_1^{(n)}\left(\M_1,Y^n\right) \;   \in\{0,1\}.
\end{equation}
Relay R$_Y$ also computes a bit-string message $\M_2 = \phi_2^{(n)}\left(Y^n,\M_1\right)$ using some encoding function $\phi_2^{(n)}: \mathcal{Y}^n\times\{0,1\}^{\star}\to\{0,1\}^{\star}$ that satisfies the expected rate constraint
\begin{equation}\label{eq:Rate2}
	\mathbb{E}\left[\mathrm{len}\left(\M_2\right)\right]\leq nR_2.
\end{equation} Then it sends $\M_2$ to the receiver R$_Z$, which guesses hypothesis $\mathcal{H}$ using   its  observation $Z^n$ and the received message $\M_2$, i.e.,  using a decision function $g_2^{(n)} : \mathcal{Z}^n \times \{0,1\}^{\star} \to \{0,1\}$, it produces the guess:
\begin{equation}
	\hat{\mathcal{H}}_Z = g_2^{(n)}\left(\M_2,Z^n\right) \;   \in\{0,1\}.
\end{equation}

The  goal is to design encoding and decision functions such that their type-I error probabilities 
\begin{IEEEeqnarray}{rCl}
	\alpha_{1,n} &\triangleq& \Pr[\hat{\mathcal{H}}_Y = 1|\mathcal{H}=0]\label{eq:type1constraint1}\\
	\alpha_{2,n} &\triangleq& \Pr[\hat{\mathcal{H}}_Z = 1|\mathcal{H}=0]\label{eq:type1constraint2}
\end{IEEEeqnarray}
stay below given thresholds $\epsilon_1 > 0$, $\epsilon_2 > 0$ and the type-II error probabilities
\begin{IEEEeqnarray}{rCl}
	\beta_{1,n} &\triangleq& \Pr[\hat{\mathcal{H}}_Y = 0|\mathcal{H}=1]\\
	\beta_{2,n} &\triangleq& \Pr[\hat{\mathcal{H}}_Z = 0|\mathcal{H}=1]
\end{IEEEeqnarray}
decay to 0 with largest possible exponential decay.  

\begin{definition} Fix maximum type-I error probabilities $\epsilon_1,\epsilon_2 \in (0,1)$ and rates $R_1,R_2 \geq 0$. The exponent pair $(\theta_1,\theta_2)$ is called \emph{$(\epsilon_1,\epsilon_2)$-achievable} if there exists a sequence of encoding and decision functions $\{\phi_1^{(n)},\phi_2^{(n)},g_1^{(n)},g_2^{(n)}\}_{n\geq 1}$ satisfying $\forall j \in \{1,2\}$:
	\begin{IEEEeqnarray}{rCl}
		\mathbb{E}[\text{len}(\M_i)] &\leq& nR_j, \label{rate_constraint} \\
		\varlimsup_{n \to \infty}\alpha_{j,n} & \leq& \epsilon_j,\label{type1constraint1}\\ 
		\label{thetaconstraint}
		\varliminf_{n \to \infty}  {1 \over n} \log{1 \over \beta_{j,n}} &\geq& \theta_j.
	\end{IEEEeqnarray}
\end{definition}
\begin{definition}
	The closure of the set of all $(\epsilon_1,\epsilon_2)$-achievable exponent pairs $(\theta_{1},\theta_{2})$ is called the \emph{$(\epsilon_1,\epsilon_2)$-exponents region} (or exponents region for short) and is denoted by $\mathcal{E}^*(R_1,R_2,\epsilon_1,\epsilon_2)$. 
	
	The maximum  exponents that are achievable at  each  of the two decision centers  are also of interest:
	\begin{IEEEeqnarray}{rCl}
		\theta^*_{1,\epsilon_1}(R_1) &:=& \max \{\theta_{1} \colon \, (\theta_{1},\theta_{2}) \in \mathcal{E}^*(R_1,R_2,\epsilon_1,\epsilon_2)  \nonumber \\
		&& \hspace{1.9cm} \textnormal{ for some } \epsilon_2>0,\theta_2 \geq 0\} \label{eq:maxtheta1}\\
		\theta^*_{2,\epsilon_2}(R_1,R_2) &:=& \max\{\theta_{2} \colon \, (\theta_{1},\theta_{2}) \in \mathcal{E}^*(R_1,R_2,\epsilon_1,\epsilon_2) \nonumber \\
		&& \hspace{1.9cm} \textnormal{ for some } \epsilon_1>0, \theta_1 \geq 0\}.\IEEEeqnarraynumspace\label{eq:maxtheta2}
	\end{IEEEeqnarray}
\end{definition}

		\section{Main Results}

	Our main result provides an exact characterization of the exponents region $\mathcal{E}^*(R_1,R_2,\epsilon_1,\epsilon_2)$.
	
	\begin{thm}\label{thm1}
		$\forall \, \epsilon_1+\epsilon_2\leq1$,   the exponents region $\mathcal{E}^*(R_1,R_2,\epsilon_1,\epsilon_2)$ \emph{is the set of} all ($\theta_{1},\theta_{2}$) pairs  satisfying 
		\begin{subequations}\label{eq:E1}
			\begin{IEEEeqnarray}{rCl}
				\theta_{1} &\leq& \min\{I(U_1;Y),I(U_2;Y)\}, \\
				\theta_{2} &\leq& \min\{I(U_2;Y)+ I(V_2;Z),I(U_3;Y)+ I(V_3;Z)\},\IEEEeqnarraynumspace
			\end{IEEEeqnarray}
			for some conditional pmfs $P_{U_1|X},P_{U_2|X},P_{U_3|X},P_{V_1|Y},P_{V_2|Y}$ and a number   $\sigma \in[1-(\epsilon_1+\epsilon_2),  1-\max\{\epsilon_1,\epsilon_2\}]$ so that
			\begin{IEEEeqnarray}{rCl}
					R_1& \geq & (1-\epsilon_1 -\sigma)I({U}_1;X) + \sigma I(U_2;X) \nonumber \\
					&& \; + (1-\epsilon_2 -\sigma)I({U}_3;X) , \label{eq:R1une} \\
				R_2& \geq & \sigma I(V_2;Y) + (1-\epsilon_2 -\sigma)I(V_3;Y). \label{eq:R2une}
			\end{IEEEeqnarray}
		\end{subequations}

	\end{thm}
	\begin{IEEEproof} Achievability is proved in Section~\ref{sec:Ach}, and the converse is proved in Section~\ref{sec:Converse}.	\end{IEEEproof}
	
	It can be shown that in the special case  $\epsilon_1 = \epsilon_2$, in Theorem~\ref{thm1} one can set without loss in optimality $\sigma = (1-\epsilon_1) = (1-\epsilon_2)$, $U_1=U_3=X$,  $V_3=Y$. This recovers the simpler characterization of the exponents region in \cite[Theorem 1]{HWS21}.  The result is presented in the following corollary, where for readability we exchanged $U_2$ by $U$ and $V_2$ by $V$.
	\begin{corollary}[Theorem 1 in \cite{HWS21}]\label{corol1}
If $\epsilon_1=\epsilon_2=\epsilon$, then the exponents region $\mathcal{E}^*(R_1,R_2,\epsilon_1,\epsilon_2)$ \emph{is the set of} all ($\theta_{1},\theta_{2}$) pairs  satisfying 
\begin{subequations}\label{eq:E2}
	\begin{IEEEeqnarray}{rCl}
		\theta_{1} &\leq& I(U;Y), \\
		\theta_{2} &\leq& I(U;Y)+ I(V;Z)\IEEEeqnarraynumspace
	\end{IEEEeqnarray}
	for some conditional pmfs $P_{U|X},P_{V|Y}$ so that
	\begin{IEEEeqnarray}{rCl}
		R_1& \geq & (1-\epsilon)I({U};X) ,\label{eq:R1equal} \\
		R_2& \geq & (1-\epsilon)I({V};Y).\label{eq:R2equal}
	\end{IEEEeqnarray}
\end{subequations}
	\end{corollary}	
\begin{IEEEproof}
See \cite{HWS21}.
\end{IEEEproof}

We remark the factors $(1-\epsilon)$ in the rate constraints  \eqref{eq:R1equal} and \eqref{eq:R2equal} compared to the optimal exponents under a maximum rate constraint determined in \cite{Vincent}. Under equal type-I error probabilities $\epsilon_1=\epsilon_2=\epsilon$, the \emph{expected} rate constraint thus allows  to boost the communication rates by a factor $(1-\epsilon)^{-1}$ compared to \emph{maximum rate constraints}. Similar boosts can also be observed  in the rate constraints \eqref{eq:R1une} and \eqref{eq:R2une} under general maximum type-I error probabilities $\epsilon_1,\epsilon_2$. 

\begin{example}
In this example, we confirm the benefit of variable-length coding compared to fixed-length coding for general permissible type-I error probabilities. 
Let $X,S,T$ be independent Bernoulli random variables of parameters $p_{X}=0.5,p_S=0.9,p_T=0.8$ and set $Y=X \oplus S$ and $Z=Y \oplus T$. We consider $\epsilon_1=0.1>\epsilon_2=0.05$ and we plot in Fig.~\ref{fig:Simulation} the optimal error exponents region $\mathcal{E}^*$ for $R_1=R_2=0.5$, which shows a tradeoff between the two exponents at the relay and the receiver. As already mentioned, such a tradeoff does not exist in the case of equal type-I error probabilities $\epsilon_1=\epsilon_2=0.05$ (obtained by Corollary~\ref{corol1}). Fig.~\ref{fig:Simulation} illustrates also the gain obtained by the expected rate constraints as opposed to the maximum rate-constraint; in fact, the rectangular region $\mathcal{E}_{\textnormal{maxR}}^*$ shows the maximum exponents region under maximum rate constraints $R_1=R_2=0.5$ for any values of $\epsilon_1,\epsilon_2$. (Under maximum rate constraints a strong converse holds, and the exponents region $\mathcal{E}_{\textnormal{maxR}}^*$ does not depend on $\epsilon_1,\epsilon_2$.)

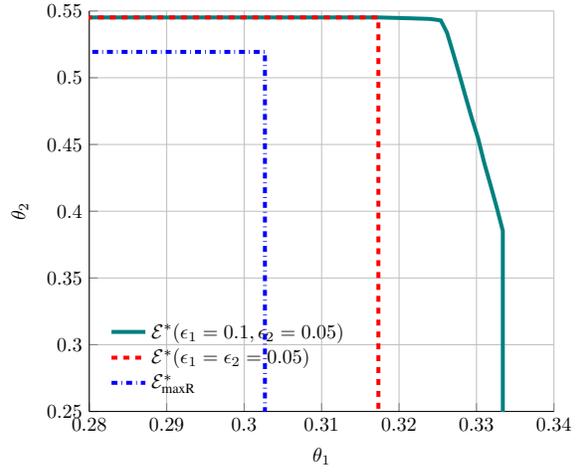
\begin{figure}[htbp]
	\begin{tikzpicture} [every pin/.style={fill=white},scale=.75]
		\begin{axis}[scale=1,
			width=.93\columnwidth,
			scale only axis,
			xmin=0.28,
			xmax=0.34,
			xmajorgrids,
			xlabel={$\theta_1$},
			ymin=0.25,
			ymax=0.55,
			ymajorgrids,
			ylabel={$\theta_2$},
			axis x line*=bottom,
			axis y line*=left,
			legend pos=south west,
			legend style={draw=none,fill=none,legend cell align=left, font=\normalsize}
			]
			\addplot[color=teal,solid,line width=2pt]
			table[row sep=crcr]{
				0 			0.545168366705219\\	
				0.317331560991042 	0.545168366705219\\
				0.318140205741108			0.545039495579276\\
				0.318948302601003			0.544909241426028\\
				0.319755849576391		0.544777610338923\\
				0.320562845046756		0.544644608781133\\
				0.321369287253390		0.544544608781133\\
				0.322175174483189			0.544374521788613\\
				0.322980504965110		0.544237449098050\\
				0.323785276846714		0.544099032683522\\
				0.324589488683081		0.543614544183753\\
				0.325393138316254			0.542872699645705\\
				0.326196224323272		0.533708136954993\\
				0.326998744561696		0.518186013401810\\
				0.327800697537541		0.502168390420586\\
				0.328602081133226			0.485574818363153\\
				0.329402893887459		0.469470785792897\\
				0.330203133683210		0.454676138053769\\
				0.331002798803622		0.436092694196009\\
				0.331801887323932			0.419822457889853\\
				0.332600397367412		0.403074210537218\\
				0.333398327353279		0.385348529799292\\
				0.333398327353279		0\\};
			\addlegendentry{$\mathcal{E}^*(\epsilon_1=0.1,\epsilon_2=0.05)$}

			\addplot[color=red,dashed,line width=2pt]
			table[row sep=crcr]{
				0					0.545168366836137\\
				0.317331560983063		0.545168366836137\\
				0.317331560983063			0\\};
			\addlegendentry{$\mathcal{E}^*(\epsilon_1=\epsilon_2=0.05)$}
		
		\addplot[color=blue,dashdotted,line width=2pt]
		table[row sep=crcr]{
			0					0.519366906171714\\
			0.302684134602472		0.519366906171714\\
			0.302684134602472			0\\};
		\addlegendentry{$\mathcal{E}_{\textnormal{maxR}}^*$}
		\end{axis} 		
	\end{tikzpicture}
	\caption{Error exponents regions under expected and maximum rate constraints when $\epsilon_1\geq\epsilon_2$. 
	}\vspace{-0.3cm}
	\label{fig:Simulation} 
\end{figure}
\end{example}

\section{General Achievability Scheme}\label{sec:Ach}
We provide a general coding and decision scheme that includes the coding and decision schemes described in \cite[Section~III]{HWS21}. The idea is to employ three different versions of the basic two-hop scheme \cite{Michele}, depending on the observed sequence $x^n$. For each version, we can  choose different codebooks, rates, and decision making strategy. To distinguish each case, 2-bit flags are added to the beginning of the messages. Details are as follows.

Fix a number $\mu  \in [0,\sigma - (1-(\epsilon_1+\epsilon_2))]$, where the interval is nonempty by the assumption $\sigma \geq 1- (\epsilon_1+\epsilon_2)$. Choose a subset $\mathcal{S}_n \subseteq \mathcal{T}_\mu^{(n)}(P_X)$ of probability 
\begin{IEEEeqnarray}{rCl}
	\mathrm{Pr}\left[X^n \in \mathcal{S}_n\right] &=& \sigma + \epsilon_1 +\epsilon_2 -1 -\mu,
\end{IEEEeqnarray}
and partition the remaining subset of $\mathcal{X}^n$ into three disjoint sets $\mathcal{D}_1$, $\mathcal{D}_2$ and $\mathcal{D}_3$
\begin{IEEEeqnarray}{rCl}
	\mathcal{D}_1 \cup \mathcal{D}_2\cup\mathcal{D}_3&=&\mathcal{X}^n \backslash \mathcal{S}_n \quad \nonumber \\
	  \mathcal{D}_i \cap \mathcal{D}_j &=&\emptyset,\quad i,j \in \{1,2,3\}, i\neq j
\end{IEEEeqnarray}
such that 
\begin{IEEEeqnarray}{rCl}
	\mathrm{Pr}\left[X^n \in \mathcal{D}_1\right] &=& 1- \epsilon_1 - \sigma\\
	\mathrm{Pr}\left[X^n \in \mathcal{D}_2\right] &=& \sigma +\mu\\
	\mathrm{Pr}\left[X^n \in \mathcal{D}_3\right] &=& 1- \epsilon_2 - \sigma.
\end{IEEEeqnarray}
We further split $R_1=R_{1,1}+R_{1,2}+R_{1,3}$ and $R_2=R_{2,2}+R_{2,3}$ for $R_{1,1},R_{1,2},R_{1,3},R_{2,2},R_{2,3}>0$.

Whenever $X^n \in \mathcal{S}_n$,  {\Tx} and {\Rel} both send the 2-bit flag $\M_1=\M_2=[0,0]$, and {\Rel} and {\Rec} declare $\hat{\mathcal{H}}_Y = \hat{\mathcal{H}}_Z = 1$. 

Whenever $X^n \in \mathcal{D}_1$, {\Tx} and {\Rel} follow the basic single-hop scheme in \cite{Ahlswede,Han} (which is included in the two-hop scheme \cite{Michele} as a special case) with  a choice of parameters $\mu, P_{U_1|X}$ satisfying 
\begin{IEEEeqnarray}{rCl}
	R_{1,1} &=&(1-\epsilon_1-\sigma)(I(U_1;X) + 2\mu),
\end{IEEEeqnarray}
and where {\Tx}  additionally sends a $[0,1]$-flag at the beginning  of $\M_1$ to \Rel, which simply relays this flag $\M_2=[0,1]$ without adding additional information. 
Upon observing $\M_2=[0,1]$, {\Rec} immediately  declares $\hat{\mathcal{H}}_Z=1$.

Whenever $X^n \in \mathcal{D}_2$,
\Tx, \Rel, and {\Rec} follow the basic two-hop scheme in \cite{Michele} but now for a different choice of parameters $\mu, P_{U_2|X}, P_{V_2|Y}$ satisfying 
\begin{IEEEeqnarray}{rCl}
	R_{1,2} &=&(\sigma+\mu)(I(U_2;X) + 2\mu)\\
	R_{2,2} &=& (\sigma+\mu)(I(V_2;Y) + 2\mu).
\end{IEEEeqnarray}

Whenever $X^n \in \mathcal{D}_3$,
\Tx, \Rel, and \Rec \, follow the basic two-hop scheme but now  for parameters $\mu, P_{U_3|X}, P_{V_3|Y}$ satisfying 
\begin{IEEEeqnarray}{rCl}
	R_{1,3} &=&(1-\epsilon_2-\sigma)(I(U_3;X) + 2\mu)\\
	R_{2,3} &=&(1-\epsilon_2-\sigma)( I(V_3;Y) + 2\mu),
\end{IEEEeqnarray}
and \Tx \,and \Rel \,add a $[1,1]$-flag to their messages  $\M_1$ and $\M_2$ to indicate to \Rel \,and \Rec \,that $X^n \in \mathcal{D}_3$. Here, we note that \Rel, upon observing the $[1,1]$-flag, declares $\hat{\mathcal{H}}_Y = 1$ even if the computed decision $\hat{\mathcal{H}}_{Y,3}$ following the basic two-hop scheme is different. 

In a similar way to \cite{HWS21}, it can be shown that this  scheme achieves the error exponents in Theorem~\ref{thm1} when $n\to \infty$ and $\mu \downarrow 0$. Details are presented in Appendix \ref{app1}.
	\section{Converse Proof to Theorem~\ref{thm1}}\label{sec:Converse}

	Fix $\theta_{1} < \theta_{1,\epsilon_1}^*(R_1)$, $\theta_{2} < \theta_{2,\epsilon_2}^*(R_1,R_2)$, a sequence (in $n$) of encoding and decision functions satisfying the constraints on the rate and the error probabilities in \eqref{rate_constraint}--\eqref{thetaconstraint}. Our proof relies on the following lemma:\vspace{2mm}
	
\begin{lemma}\label{lem:receiverconverse}
Fix a blocklength $n$ and a set $\mathcal{D}\subseteq \mathcal{X}^n\times\mathcal{Y}^n$ of positive probability, and let  the tuple ($\tilde{\M}_1,\tilde{\M}_2,\tilde{X}^n,\tilde{Y}^n,\tilde{Z}^n$)  follow the pmf 
\begin{IEEEeqnarray}{rCl}
	\lefteqn{P_{{\tilde{\M}_1}{\tilde{\M}_2}\tilde{X}^n\tilde{Y}^n\tilde{Z}^n}(\m_1,\m_2,x^n,y^n,z^n) \triangleq} \qquad \nonumber \\
	&& P_{X^nY^nZ^n}(x^n,y^n,z^n)\cdot{\mathbbm{1} \{(x^n,y^n)\in \mathcal{D}\} \over P_{X^nY^n}(\mathcal{D})}\nonumber\\ && \qquad \cdot{\mathbbm{1}\{\phi_1(x^n)=\m_1\}}
	\cdot{\mathbbm{1}\{\phi_2(y^n,\phi_1(x^n))=\m_2\}}. \IEEEeqnarraynumspace \label{pmftildedoubleprime2_lemma}
\end{IEEEeqnarray}
Further, define ${U} \triangleq (\tilde{\M}_1,\tilde{X}^{T-1}\tilde{Y}^{T-1},T)$, ${V} \triangleq (\tilde{\M}_2,\tilde{X}^{T-1}\tilde{Y}^{T-1},T)$, $\tilde{X} \triangleq\tilde{X}_T$, $\tilde{Y} \triangleq \tilde{Y}_T$, and $\tilde{Z}\triangleq \tilde{Z}_T$,  where $T$ is uniform over $\{1,\ldots,n\}$ and independent of all other random variables, and notice the Markov chain $V \to \tilde{Y}\to \tilde{Z}$.
The following (in)equalities hold:
\begin{IEEEeqnarray}{rCl}
	H(\tilde{M}_1) &\geq& nI(U;\tilde{X}) + \log P_{X^nY^n}(\mathcal{D}),\\
	H(\tilde{M}_2) &\geq& nI(V;\tilde{Y}) + \log P_{X^nY^n}(\mathcal{D}),\\
I(U;\tilde{Y}|\tilde{X}) &= &\o_1(n), 
\end{IEEEeqnarray}
where $\o_1(n)$ is a function that tends to 0 as $n\to \infty$.

Let $\eta > 0$ be arbitrary. If
\begin{equation}\label{eq:lemma1_cond1}
	\Pr[\hat{\mathcal{H}}_Z=0|\mathcal{H}=0,X^n=x^n,Y^n=y^n]\geq \eta, \;\; \forall (x^n,y^n) \in \mathcal{D},
\end{equation}
then
\begin{IEEEeqnarray}{rCl}
	\lefteqn{-{1\over n}\log\Pr[\hat{\mathcal{H}}_Z = 0| \mathcal{H}=1,(X^n,Y^n) \in \mathcal{D}]} \qquad \qquad \qquad \nonumber\\
	&& \leq I(U;\tilde{Y}) + I(V;\tilde{Z}) + \o_2(n),
\end{IEEEeqnarray}
and if
\begin{equation}\label{lem:cond2}
	\Pr[\hat{\mathcal{H}}_Y=0|\mathcal{H}=0,X^n=x^n,Y^n=y^n] \geq \eta, \;\; \forall (x^n,y^n) \in \mathcal{D},
\end{equation}
then
\begin{equation}
-{1\over n}\log\Pr[\hat{\mathcal{H}}_Y = 0| \mathcal{H}=1,(X^n,Y^n)\in\mathcal{D}] \leq I(U;\tilde{Y}) + \o_3(n),
\end{equation}
where $\o_2(n),\o_3(n)$ are functions that tend to $0$ as $n \to \infty$.
\end{lemma}
\begin{IEEEproof}
See Appendix~\ref{app_lemma1}.
\end{IEEEproof}

We now prove the converse to Theorem~\ref{thm1}. Fix a positive number $\eta >0$. Denote   for each blocklength $n$, the set of strongly jointly typical sequences in $\mathcal{X}^n\times\mathcal{Y}^n$ by $\mathcal{T}_{\mu_n}^{(n)}(P_{XY})$  and set $\mu_n=n^{-1/3}$. Define the sets
	\begin{IEEEeqnarray}{rCl}\label{Bn}
		\mathcal{B}_{1}(\eta) &\triangleq& \{(x^n,y^n) \in \mathcal{T}_{\mu_n}^{(n)}(P_{XY}) \colon \nonumber \\ && \;\; \mathrm{Pr}[\hat{\mathcal{H}}_Y=0 | X^n = x^n, Y^n=y^n, \mathcal{H}=0] \geq \eta\}, \IEEEeqnarraynumspace
		\label{eq:Bn1def}\\
		\mathcal{B}_{2}(\eta) &\triangleq& \{(x^n,y^n) \in \mathcal{T}_{\mu_n}^{(n)}(P_{XY}) \colon \nonumber \\ && \;\; \mathrm{Pr}[\hat{\mathcal{H}}_Z=0 | X^n = x^n, Y^n=y^n, \mathcal{H}=0] \geq \eta\},\IEEEeqnarraynumspace
		  \label{eq:Bn2def}\\
		\mathcal{D}_{2}(\eta) &\triangleq& 	\mathcal{B}_{1}(\eta) \cap 	\mathcal{B}_{2}(\eta),\\
		\mathcal{D}_{1}(\eta) &\triangleq& \mathcal{B}_{1}(\eta) \backslash \mathcal{D}_{2}(\eta), \label{D1n}\\
		\mathcal{D}_{3}(\eta) &\triangleq& \mathcal{B}_{2}(\eta) \backslash \mathcal{D}_{2}(\eta). \label{D2n} 
	\end{IEEEeqnarray}
Further define for each $n$ the probabilities
\begin{IEEEeqnarray}{rCl}
	\Delta_i &\triangleq& P_{X^nY^n}(\mathcal{D}_{i}(\eta)), \quad i\in\{1,2,3\},
\end{IEEEeqnarray}
and notice that 
\begin{equation}\label{eq:sum}
\Delta_1 +\Delta_2 = P_{X^nY^n}(\mathcal{B}_{1}(\eta)) \;\; \textnormal{and} \;\; \Delta_2 +\Delta_3 = P_{X^nY^n}(\mathcal{B}_{2}(\eta)),
\end{equation}
where by  \cite[Remark to Lemma~2.12]{Csiszarbook} and the type-I error probability constraints in \eqref{type1constraint1}:
	\begin{IEEEeqnarray}{rCl}\label{eq:PB}
		P_{X^nY^n}(\mathcal{B}_{j}(\eta)) &\geq& {1 - \epsilon_j - \eta \over{1 - \eta}} - {\vert{\mathcal{X}}\vert \vert{\mathcal{Y}}\vert \over{4 \mu_n^2 n}}, \quad j\in\{1,2\}. \IEEEeqnarraynumspace
		\end{IEEEeqnarray}

For any $i\in\{1,2,3\}$ such that   $\Delta_i > 0$, we apply Lemma~\ref{lem:receiverconverse} to the set $\mathcal{D}_i$. This allows to conclude that for any $i\in \{1,2,3\}$ with  $\Delta_i > 0$ there exists a pair $U_i,V_i$ satisfying the Markov chain $V_i \to \tilde{Y} \to \tilde{Z}$ and the (in)equalities 
\begin{IEEEeqnarray}{rCl}
	H(\tilde{M}_{1,i}) &\geq& nI(U_i;\tilde{X}_i) + \log P_{X^nY^n}(\mathcal{D}_i),\; i\in\{1,2,3\}, \label{eq:M1i} \IEEEeqnarraynumspace\\
	H(\tilde{M}_{2,i}) &\geq& nI(V_{i};\tilde{Y}_{i}) + \log P_{X^nY^n}(\mathcal{D}_i), \;\; i \in \{2,3\},\label{eq:M2i}\\
	I(U_i; \tilde{Y}|\tilde{X}) & = & \o_{1,i}(n),
\end{IEEEeqnarray}
and 
\begin{IEEEeqnarray}{rCl}
\lefteqn{	-{1\over n}\log\Pr[\hat{\mathcal{H}}_Y = 0| \mathcal{H}=1,(X^n,Y^n)\in\mathcal{D}_i]} \qquad \quad \nonumber \\
	 && \leq I(U_i;\tilde{Y}_i) + \o_{3,i}(n), \qquad \qquad \qquad i\in\{1,2\},\IEEEeqnarraynumspace
\end{IEEEeqnarray}
\begin{IEEEeqnarray}{rCl}
	\lefteqn{-{1\over n}\log\Pr[\hat{\mathcal{H}}_Z = 0| \mathcal{H}=1, (X^n,Y^n) \in \mathcal{D}_i]} \qquad \quad \nonumber\\
	&& \leq I(U_i;\tilde{Y}_i) + I(V_i;\tilde{Z}_i) + \o_{2,i}(n), \quad i\in\{2,3\}, \label{eq:sumI} \IEEEeqnarraynumspace
\end{IEEEeqnarray}
where for each $i$  the functions $\o_{1,i}(n), \o_{2,i}(n),\o_{3,i}(n)\to 0$ as $n\to \infty$ and the random variables $\tilde{X}_{i},\tilde{Y}_i, \tilde{Z}_i, \tilde{M}_{1,i},\tilde{M}_{2,i}$ are defined as in the lemma applied to the subset $\mathcal{D}_i$. 

Notice that when  $\Delta_i = 0$, the trivial choice  $(U_i=\tilde{X}, V_i=\tilde{Y})$ satisfies the Markov chain $V_i \to \tilde{Y} \to \tilde{Z}$ and  Inequalities \eqref{eq:M1i}--\eqref{eq:sumI}. The existence of the desired pair $(U_i,V_i)$ as described in the previous paragraph can thus  be concluded for any value of $i\in\{1,2,3\}$.

We continue with  the total law of probability to obtain:
		\begin{IEEEeqnarray}{rCl}
	-\frac{1}{n} \log \beta_{1,n}  &\leq& \min\{I(U_1;\tilde{Y});I(U_2;\tilde{Y})\} + \o_3(n),\label{eq:R1thetaslasteq}\\
		-\frac{1}{n} \log \beta_{2,n} &\leq& \min\{I(U_2;\tilde{Y})+I(V_2;\tilde{Z});\nonumber \\ 
		&& \qquad \quad I(U_3;\tilde{Y}) +I(V_3;\tilde{Z})\} +\o_2(n),\label{eq:R1R2thetaslasteq}\IEEEeqnarraynumspace
	\end{IEEEeqnarray}
	where $\o_2(n)$ and $\o_3(n)$ are functions tending to 0 as $n\to \infty$.

Further define the following random variables for $j\in\{1,2\}$ and $i\in\{1,2,3\}$
	\begin{equation}
		{\tilde{L}_{j,i}} \triangleq \mathrm{len}({\tilde{\M}_{j,i}}),
	\end{equation}
	By the rate constraints  \eqref{eq:Rate1} and \eqref{eq:Rate2}, and the definition of the random variables $\tilde{\M}_{j,i}$, we obtain by the total law of expectations:
	\begin{IEEEeqnarray}{rCl}
		nR_j &\geq& \mathbb{E}[L_j]\\
		&\geq&\sum_{i\in\{1,2,3\}}\mathbb{E}[\tilde{L}_{j,i}]\Delta_i. \label{ELi}
	\end{IEEEeqnarray} 
Moreover,  
	\begin{IEEEeqnarray}{rCl}
		H(\tilde{\M}_{j,i}) &=& H(\tilde{\M}_{j,i},\tilde{L}_{j,i})\\
		&=& \sum_{l_i} \Pr[\tilde{L}_{j,i} = l_i]H(\tilde{\M}_{j,i}|\tilde{L}_{j,i}=l_i) + H(\tilde{L}_{j,i})\IEEEeqnarraynumspace\\
		&\leq& \sum_{l_i} \Pr[\tilde{L}_{j,i} = l_i]l_i + H(\tilde{L}_{j,i}) \label{HMi_ineq1}\\
		&=& \mathbb{E}[\tilde{L}_{j,i}] + H(\tilde{L}_{j,i}),
	\end{IEEEeqnarray}
which combined with \eqref{ELi} establishes
\begin{IEEEeqnarray}{rCl}
\hspace{-2mm}	\sum_{i\in\{1,2,3\}} \Delta_i H(\tilde{\M}_{1,i}) &\leq& \sum_{i\in\{1,2,3\}} \Delta_i\mathbb{E}[\tilde{L}_{1,i}] +  \Delta_i H(\tilde{L}_{1,i}) \IEEEeqnarraynumspace\\
		&\leq & {nR_1} \left(1 +\sum_{i\in\{1,2,3\}} h_b\left({\Delta_i \over nR_1}\right)\right), \label{M1ub'}
\end{IEEEeqnarray}
where \eqref{M1ub'} holds by \eqref{ELi} and because the entropy of the discrete and positive random variable $\tilde{L}_{1,i}$ of  mean   $\E{[\tilde{L}_{1,i}]} \leq {nR_1\over \Delta_i}$ is bounded by $ \frac{n R_1}{\Delta_i} \cdot h_b\left({\Delta_i \over nR_1}\right)$, see \cite[Theorem 12.1.1]{cover}. 

In a similar way we obtain 
\begin{IEEEeqnarray}{rCl}
	\sum_{i\in\{2,3\}}\Delta_i H(\tilde{\M}_{2,i}) \leq {nR_2} \left(1 +\sum_{i\in\{2,3\}} h_b\left({\Delta_i \over nR_2}\right)\right). \IEEEeqnarraynumspace \label{M2ub'}
\end{IEEEeqnarray}
Then by combining  \eqref{M1ub'} and \eqref{M2ub'} with \eqref{eq:M1i} and \eqref{eq:M2i}, noting \eqref{eq:sum} and \eqref{eq:PB}, and considering also \eqref{eq:R1thetaslasteq} and \eqref{eq:R1R2thetaslasteq}, we have proved so far that for all $n\geq 1$ there exist joint pmfs $P_{U_i\tilde{X}\tilde{Y}}=P_{U_i|\tilde{X}}P_{\tilde{X}\tilde{Y}}$ (abbreviated as $P_{U,i}^{(n)}$) for $i\in\{1,2,3\}$, and $P_{V_i\tilde{Y}\tilde{Z}}=P_{V_i|\tilde{Y}}P_{\tilde{Y}\tilde{Z}} $ (abbreviated as $P_{V,i}^{(n)}$) for $i\in\{2,3\}$ so that the following conditions hold (where $I_{P}$ indicates that the mutual information should be calculated according to a pmf $P$):
\begin{subequations}\label{eq:conditions}
	\begin{IEEEeqnarray}{rCl}			
		R_1 &\geq  &\sum_{i\in\{1,2,3\}}\big(I_{P_{U,i}^{(n)}}({U}_i;\tilde{X}) + g_{1,i}(n)\big)\cdot g_{2,i}(n,\eta) ,\label{eq:R111} \IEEEeqnarraynumspace \\
		R_2 &\geq  &\sum_{i\in\{2,3\}}\big(I_{P_{V,i}^{(n)}}({V}_i;\tilde{Y}) + g_{1,i}(n)\big)\cdot g_{2,i}(n,\eta), \label{eq:R222}\\
		\theta_1 &\leq& \min \{I_{P_{U,1}^{(n)}}({U}_1;\tilde{Y}), I_{P_{U,2}^{(n)}}({U}_2;\tilde{Y})\} + g_{3,1}(n) \label{eq:theta111}\IEEEeqnarraynumspace\\
		\theta_2 &\leq& \min \{\big(I_{P_{U,2}^{(n)}}({U}_2;\tilde{Y}) + I_{P_{V,2}^{(n)}}(V_2;\tilde{Z})\big) + g_{3,2}(n),\nonumber\\
		&&\quad \big(I_{P_{U,3}^{(n)}}({U}_3;\tilde{Y}) + I_{P_{V,3}^{(n)}}(V_3;\tilde{Z})\big) + g_{3,3}(n)\}  \label{eq:theta222}\IEEEeqnarraynumspace\\
		g_{4,i}(n) &=& I_{P_{U,i}^{(n)}}(\tilde{Y};{U}_i|\tilde{X}) , \quad i \in \{1,2,3\}\label{eq:last_cond}
	\end{IEEEeqnarray}
\end{subequations}
for some functions $g_{1,i}(n), g_{2,i}(n,\eta), g_{3,i}(n), g_{4,i}(n)$ with the following asymptotic behaviors:
\begin{IEEEeqnarray}{rCl}
	\lim_{n\to \infty} g_{1,i}(n) &=& 0, \qquad  \forall i \in \{1,2,3\},\IEEEeqnarraynumspace\\ 
	\lim_{n\to \infty} g_{3,i}(n) &=& 0, \qquad  \forall i \in \{1,2,3\},\\
	\lim_{n\to \infty} g_{4,i}(n)& = & 0, \qquad  \forall i \in \{1,2,3\},\\
	\lim_{n\to \infty} \left(g_{2,1}(n,\eta)+g_{2,2}(n,\eta)\right) & =  &\frac{1-\epsilon_1-\eta}{1-\eta},\\
	\lim_{n\to \infty} \left(g_{2,2}(n,\eta)+g_{2,3}(n,\eta)\right) & =  &\frac{1-\epsilon_2-\eta}{1-\eta}. \IEEEeqnarraynumspace
\end{IEEEeqnarray}

The proof is  concluded by letting $n\to \infty$ and $\eta\downarrow 0$, and noting that by \eqref{eq:last_cond} the limiting pmf of the sequence $P_{U,i}^{(n)}$ satisfies the Markov condition ${U}_i \to \tilde{X}\to \tilde{Y}$. More precisely, we first observe that by  Carath\'eodory's theorem \cite[Appendix C]{ElGamal} for each $n$ there must exist random variables ${U}_1,U_2,U_3,V_2,V_3$ satisfying \eqref{eq:conditions} over  alphabets of sizes
\begin{align}
	\vert {\mathcal{U}}_i\vert &\leq \vert \mathcal{X}\vert\cdot\vert \mathcal{Y}\vert + 2, \qquad \quad i \in \{1,2,3\},\\
	\vert {\mathcal{V}}_i \vert &\leq \vert {\mathcal{U}}_i \vert\cdot|\mathcal{X}|\cdot\vert \mathcal{Y}\vert + 1, \quad i \in \{2,3\}.
\end{align}
Then we invoke the Bolzano-Weierstrass theorem and  consider for each $i \in \{1,2,3\}$ a sub-sequence  $P_{\tilde{X}\tilde{Y}{U}_i}^{(n_k)}$ that converges to a limiting pmf $P_{XYU_i}^{*}$, and for each $i \in \{2,3\}$, a sub-sequence  $P_{\tilde{Y}\tilde{Z}{V}_i}^{(n_k)}$ that converges to a limiting pmf $P_{YZV_i}^{*}$. For these limiting pmfs, which we abbreviate by $P_{U,i}^*$ and $P_{V,i}^*$ respectively, we conclude by \eqref{eq:R111}--\eqref{eq:theta222}:
\begin{IEEEeqnarray}{rCl}			
	R_1& \geq &(1-\epsilon_1 - \sigma)I_{P_{U,1}^{*}}({U}_1;{X}) + \sigma I_{P_{U,2}^{*}}({U}_2;{X})\nonumber\\
	&& \quad +  (1-\epsilon_2 - \sigma)I_{P_{U,3}^{*}}({U}_3;{X})\label{eq:R_1_f} \\
	R_2 &\geq  &\sigma I_{P_{V,2}^{*}}({V}_2;{Y}) +  (1-\epsilon_2 - \sigma)I_{P_{V,3}^{*}}({V}_3;{Y}) \\
	\theta_1 &\leq & \min \{I_{P_{U_1}^{*}}({U}_1;{Y}),I_{P_{U_2}^{*}}({U}_2;{Y})\},\label{theta_1_f}\\
	\theta_2 &\leq &\min \{I_{P_{U_2}^{*}}({U}_2;{Y})+I_{P_{V_2}^{*}}({V}_2;{Z}),\nonumber \\ 
	&& \qquad \quad I_{P_{U_3}^{*}}({U}_3;{Y})+I_{P_{V_3}^{*}}({V}_3;{Z})\},\label{theta_2_f}\end{IEEEeqnarray}
where $\sigma := \lim\limits_{n \to \infty,\eta\downarrow 0}\Delta_2$, and where due to \eqref{eq:sum}, $\sigma$ can be upper-bounded by $1-\max\{\epsilon_{1},\epsilon_{2}\}$. Furthermore, given that $\epsilon_1+\epsilon_{2}\leq 1$, then $\sigma$ can be lower-bounded by $1-(\epsilon_{1}+\epsilon_{2})$.
Notice further that since for any $k$ the pair $(\tilde{X}^{n_k},\tilde{Y}^{n_k})$ lies in the jointly typical set $\mathcal{T}^{(n_k)}_{\mu_{n_k}}(P_{XY})$, we have  $\vert P_{\tilde{X}\tilde{Y}} - P_{XY}\vert \leq \mu_{n_k}$ and thus the limiting pmfs satisfy $P^*_{XY}=P_{XY}$. Moreover, since for each $n_k$ the random variable $\tilde{Z}$ is drawn according to $P_{Z|Y}$ given $\tilde{Y}$, irrespective of $\tilde{X}$, the limiting pmf also satisfies $P_{Z|XY}^*=P_{Z|Y}$. 
We also notice $\forall i \in \{2,3\}$ that  under $P_{V_iYZ}^*$ the Markov chain
\begin{IEEEeqnarray}{rCl}\label{eq:MC2_1}
	V_i\to Y \to Z, 
\end{IEEEeqnarray}	
holds because  ${V}_i\to \tilde{Y} \to \tilde{Z}$  forms a Markov chain for any $n_k$. Finally, by continuity considerations and by \eqref{eq:last_cond}, the following Markov chain must hold under $P_{U_iXY}^*$ $\forall i \in \{1,2,3\}$:
\begin{IEEEeqnarray}{rCl}\label{eq:MC2_2}
	U_i \to X \to Y.
\end{IEEEeqnarray}
This concludes our converse proof.

	\section*{Acknowledgment}
	{M. Wigger and M. Hamad acknowledge funding support from the ERC under grant agreement 715111.}
	\vspace{-0.1cm}

	\appendices

\section{Analysis of the coding scheme in Section~\ref{sec:Ach}}\label{app1}
Let ${\tilde{\mathcal{H}}}_{Y,1}$ denote the hypothesis guessed by {\Rel} for the basic single-hop scheme with the first parameter choices $\mu, P_{U_1|X}$. Then let  ${\tilde{\mathcal{H}}_{Y,2}}$ and  $\tilde{\mathcal{H}}_{Z,2}$ denote the hypotheses guessed by {\Rel} and {\Rec} for the basic two-hop scheme with the second parameter choices $\mu, P_{U_2|X}, P_{V_2|Y}$. Similarly, let ${\tilde{\mathcal{H}}}_{Z,3}$ be the hypothesis  produced by {\Rec} for the basic two-hop scheme with the third parameter choices $\mu, P_{U_3|X}, P_{V_3|Y}$. 
We obtain for the type-I error probabilities:  
\begin{IEEEeqnarray}{rCl}
	\alpha_{1,n} &=& \Pr[\hat{\mathcal{H}}_Y = 1, X^n \in (\mathcal{S}_{n}\cup\mathcal{D}_3) |\mathcal{H} = 0] \nonumber \\
	&& + \Pr[\hat{\mathcal{H}}_Y = 1, X^n \in \mathcal{D}_{1}|\mathcal{H} = 0] \nonumber \\
	&& + \Pr[\hat{\mathcal{H}}_Y = 1, X^n \in \mathcal{D}_{2}|\mathcal{H} = 0]\IEEEeqnarraynumspace\\
	&=& \Pr[X^n \in (\mathcal{S}_{n}\cup\mathcal{D}_3)|\mathcal{H}=0] \nonumber\\
	&& +  \Pr[\tilde{\mathcal{H}}_{Y,1} = 1, X^n \in \mathcal{D}_{1}|\mathcal{H}=0]  \nonumber \\
	&& +  \Pr[\tilde{\mathcal{H}}_{Y,2} = 1, X^n \in \mathcal{D}_{2}|\mathcal{H}=0] \IEEEeqnarraynumspace\\
	&\leq& \epsilon_1 - \mu + \Pr[\tilde{\mathcal{H}}_{Y,1} = 1|\mathcal{H}=0] \nonumber \\ 
	&& +  \Pr[\tilde{\mathcal{H}}_{Y,2} = 1|\mathcal{H}=0]
\end{IEEEeqnarray}
and
\begin{IEEEeqnarray}{rCl}
	\alpha_{2,n} &=& \Pr[\hat{\mathcal{H}}_Z = 1, X^n \in (\mathcal{S}_{n}\cup\mathcal{D}_1) |\mathcal{H} = 0] \nonumber \\ 
	&& + \Pr[\hat{\mathcal{H}}_Z = 1, X^n \in \mathcal{D}_{2} |\mathcal{H} = 0] \nonumber \\ 
	&& +  \Pr[\hat{\mathcal{H}}_Z = 1, X^n \in \mathcal{D}_3|\mathcal{H} = 0]\IEEEeqnarraynumspace\\
	&=& \Pr[X^n \in (\mathcal{S}_{n}\cup\mathcal{D}_1)|\mathcal{H}=0] \nonumber \\ 
	&& +  \Pr[\tilde{\mathcal{H}}_{Z,2} = 1, X^n \in \mathcal{D}_2|\mathcal{H}=0] \nonumber \\
	&& + \Pr[\tilde{\mathcal{H}}_{Z,3} = 1, X^n \in \mathcal{D}_{3}|\mathcal{H}=0] \IEEEeqnarraynumspace\\
	&\leq& \epsilon_2 - \mu + \Pr[\tilde{\mathcal{H}}_{Z,2} = 1|\mathcal{H}=0] \nonumber\\ && + \Pr[\tilde{\mathcal{H}}_{Z,3} = 1|\mathcal{H}=0].
\end{IEEEeqnarray}
Since by \cite{Michele,JSAIT}, $\forall i \in \{1,2\}$ and $\forall j \in \{2,3\}$ the probabilities $\Pr[\tilde{\mathcal{H}}_{Y,i} = 1|\mathcal{H}=0]$, and $ \Pr[\tilde{\mathcal{H}}_{Z,j} = 1|\mathcal{H}=0]$ tend to 0 as $n \to \infty$, we conclude that for the general scheme in Section~\ref{sec:Ach} $\varlimsup_{n\to\infty}\alpha_{1,n} \leq \epsilon_1$ and $\varlimsup_{n\to\infty}\alpha_{2,n} \leq \epsilon_2$.

For the type-II error probabilities we obtain
\begin{IEEEeqnarray}{rCl}
	\beta_{1,n} 
	&=& \Pr[\tilde{\mathcal{H}}_{Y,1}=0, X^n\in \mathcal{D}_{1} |\mathcal{H}=1] \nonumber \\
	&& + \Pr[\tilde{\mathcal{H}}_{Y,2}=0, X^n\in \mathcal{D}_{2} |\mathcal{H}=1]\\
	&\leq& \Pr[\tilde{\mathcal{H}}_{Y,1}=0|\mathcal{H}=1] + \Pr[\tilde{\mathcal{H}}_{Y,2}=0|\mathcal{H}=1]\IEEEeqnarraynumspace\\
	&\leq& 2^{-n\left(I(U_1;Y)+ \delta(\mu)\right)} + 2^{-n\left(I(U_2;Y)+ \delta(\mu)\right)}, \label{BetaZLAPP3}
\end{IEEEeqnarray}\vspace{-2mm}
and 
\begin{IEEEeqnarray}{rCl}
	\beta_{2,n} 
	&=& \Pr[\tilde{\mathcal{H}}_{Z,2}=0, X^n\in \mathcal{D}_2 |\mathcal{H}=1] \nonumber \\
	&& + \Pr[\tilde{\mathcal{H}}_{Z,3}=0, X^n\in \mathcal{D}_{3}|\mathcal{H}=1]\\
	&\leq& \Pr[\tilde{\mathcal{H}}_{Z,2}=0|\mathcal{H}=1] + \Pr[\tilde{\mathcal{H}}_{Z,3}=0|\mathcal{H}=1]\IEEEeqnarraynumspace\\
	&\leq& 2^{-n\left(I(U_2;Y)+I(V_2;Z) + \delta(\mu)\right)} \nonumber \\ 
	&& + 2^{-n\left(I(U_3;Y)+I(V_3;Z) + \delta(\mu)\right)}. \IEEEeqnarraynumspace  \label{BetaSWWAPP3}
\end{IEEEeqnarray}
where \eqref{BetaZLAPP3} and \eqref{BetaSWWAPP3} are proved in \cite{Michele}, and $\delta(\mu) \downarrow 0$ as $\mu \downarrow 0$. 

Notice further that for sufficiently large blocklengths $n$ that  satisfy $(2-\epsilon_1 - \epsilon_{2} -\sigma +\mu)n\mu \geq 2$ and $(1 - \epsilon_{2} +\mu)n\mu \geq 2$, the  described scheme satisfies both rate constraints:
\begin{IEEEeqnarray}{rCl}
	\E[\len(\M_1)]& \leq& (\sigma + \epsilon_{1} + \epsilon_2-1 - \mu)\cdot2 \nonumber \\ && +(1-\epsilon_1 - \sigma)\cdot (n( I(U_1;X) + \mu)+2) \nonumber \\
	&& + ( \sigma + \mu  )\cdot (n(I(U_2;X) + \mu) +2)\\
	&& + ( 1-\epsilon_2- \sigma  )\cdot (n(I(U_3;X) + \mu) +2)\IEEEeqnarraynumspace\\
	& \leq & n (R_1'+R_1''+R_1''')=nR_1 \label{eq:n_large_enough_1}
\end{IEEEeqnarray}
and 
\begin{IEEEeqnarray}{rCl}
	\E[\len(\M_2)]& \leq & (\epsilon_2-\mu)\cdot2\nonumber \\
	&&+(\sigma +\mu)\cdot (n( I(V_2;Y) + \mu)+2) \nonumber \\
	&& + ( 1 -\epsilon_2-\sigma ) \cdot (n(I(V_3;Y) + \mu)  +2) \IEEEeqnarraynumspace\\
	& \leq & n (R_2''+R_2''')=nR_2. \label{eq:n_large_enough_2}
\end{IEEEeqnarray}

Letting first $n \to \infty$ and then  $\mu \downarrow 0$, establishes the desired achievability result in \eqref{eq:E1}.

	\section{Proof of Lemma~\ref{lem:receiverconverse}}\label{app_lemma1}
Throughout this section, let $h_{b}(\cdot)$ denote the binary entropy function, and $D(P\|Q)$  the Kullback-Leibler divergence between two probability mass functions on the same alphabet.
	Note first that by \eqref{pmftildedoubleprime2_lemma}:
	\begin{equation}\label{tildedivergencerelation2_lemma}
		D(P_{\tilde{X}^n\tilde{Y}^n}||P_{XY}^{n}) \leq \log{\Delta_n^{-1}},
	\end{equation}
	where we defined $\Delta_n \triangleq P_{X^nY^n}(\mathcal{D})$.

Further define  $\tilde{V}_t\triangleq(\tilde{\M}_2,\tilde{X}^{t-1},\tilde{Y}^{t-1})$ and $\tilde{U}_t\triangleq(\tilde{\M}_1,\tilde{X}^{t-1},\tilde{Y}^{t-1})$ and notice:
	\begin{IEEEeqnarray}{rCl}
		H(\tilde{\M}_1)	&\geq& I(\tilde{\M}_1;\tilde{X}^n\tilde{Y}^n) + D(P_{\tilde{X}^n\tilde{Y}^n}||P_{XY}^n) + \log\Delta_{n}\IEEEeqnarraynumspace\label{m1entropylbstep1_lemma}\\
		&=& H(\tilde{X}^n\tilde{Y}^n) + D(P_{\tilde{X}^n\tilde{Y}^n}||P_{XY}^n) \nonumber \\ && -  H(\tilde{X}^n\tilde{Y}^n|\tilde{\M}_1) + \log\Delta_{n}\IEEEeqnarraynumspace\\
		&\geq& n [H(\tilde{X}_{T}\tilde{Y}_{T}) + D(P_{\tilde{X}_{T}\tilde{Y}_{T}}||P_{XY})] \nonumber \\ && - \sum_{t=1}^{n} H(\tilde{X}_{t}\tilde{Y}_{t}|\tilde{U}_{t})+ \log\Delta_{n}\IEEEeqnarraynumspace\label{m1entropylbstep4_lemma}\\
		&=& n [H(\tilde{X}_{T}\tilde{Y}_{T}) + D(P_{\tilde{X}_{T}\tilde{Y}_{T}}||P_{XY}) \nonumber \\ && - H(\tilde{X}_{T}\tilde{Y}_{T}|\tilde{U}_{T},T)]+ \log\Delta_{n} \label{Tuniformdef_lemma}\\
				&\geq & n [H(\tilde{X}_{T}\tilde{Y}_{T})  - H(\tilde{X}_{T}\tilde{Y}_{T}|\tilde{U}_{T},T)]+ \log\Delta_{n} \IEEEeqnarraynumspace\\
		&=& n [I(\tilde{X}\tilde{Y};U)] +  \log{\Delta_{n}} \label{eq:HM1_LB_lemma_last_eq} \\
		&\geq& n \left[I(\tilde{X};U) + {1 \over n} \log{\Delta_{n}}\right] .\IEEEeqnarraynumspace \label{eq:HM1_LB_lemma}
	\end{IEEEeqnarray}
	Here, (\ref{m1entropylbstep1_lemma}) holds by (\ref{tildedivergencerelation2_lemma}); (\ref{m1entropylbstep4_lemma}) holds by the super-additivity property in \cite[Proposition 1]{tyagi2019strong}, by the chain rule, and by the definition of $\tilde{U}_{t}$; \eqref{Tuniformdef_lemma} by defining $T$ uniform over $\{1,\dots,n\}$ independent of all other random variables; and \eqref{eq:HM1_LB_lemma_last_eq} by the definitions of $U,\tilde{X},\tilde{Y}$ in the lemma.
	
	We can lower bound the entropy of $\tilde{\M}_2$ in a similar way to obtain:
	\begin{IEEEeqnarray}{rCl}
		H(\tilde{\M}_2) &\geq& n \left[I(\tilde{Y};V) + {1 \over n} \log{\Delta_{n}}\right]. 
	\end{IEEEeqnarray}

	We next  upper bound the error exponent at the receiver. 
	
	Define
	\begin{equation}\label{eq:yz_acceptance}
		\mathcal{A}_{Z,n}(\m_2) \triangleq \{z^n \colon g_2(\m_2,z^n) = 0\},
	\end{equation}
	and its Hamming neighborhood:
	\begin{equation}
		\hat{\mathcal{A}}_{Z,n}^{\ell_n}(\m_2) \triangleq \{\tilde{z}^n : \exists \, z^n \in \mathcal{A}_{Z,n}(\m_2) \textnormal{ s.t.} \; d_H(z^n,\tilde{z}^n)\leq\ell_n\}
	\end{equation}
	for some real number $\ell_n$ satisfying $\lim_{n \rightarrow \infty} {\ell_n/n} =0 $ and $\lim_{n \to \infty} {\ell_n/\sqrt{n}} =\infty $.	
	Since by Condition \eqref{eq:lemma1_cond1}, 
	\begin{equation}\label{blowupcond2_lemma}
		P_{\tilde{Z}^n|\tilde{X}^n\tilde{Y}^n}(\mathcal{A}_{Z,n}(\m_2)|x^n,y^n) \geq \eta , \quad \forall (x^n,y^n) \in \mathcal{D},
	\end{equation}
	 the blowing-up lemma \cite{MartonBU} yields
	\begin{equation}\label{blowup2_lemma}
		P_{\tilde{Z}^n|\tilde{X}^n\tilde{Y}^n}(\hat{\mathcal{A}}_{Z,n}^{\ell_n}(\m_2)|x^n,y^n) \geq 1 - \zeta_n, \quad \forall (x^n,y^n) \in \mathcal{D},
	\end{equation}
	for a real number $\zeta_n > 0$ such that $\lim\limits_{n \to \infty} \zeta_n = 0$.\\
Define
			\begin{equation}
				{\mathcal{A}}_{Z,n} \triangleq \bigcup\limits_{\m_2 \in \mathcal{M}_2} \{\m_2\} \times {\mathcal{A}}_{Z,n}(\m_2),
			\end{equation}
		\begin{equation}
			\hat{\mathcal{A}}_{Z,n}^{\ell_n} \triangleq \bigcup\limits_{\m_2 \in \mathcal{M}_2} \{\m_2\} \times \hat{\mathcal{A}}_{Z,n}^{\ell_n}(\m_2),
	\end{equation} 
and notice that
\begin{IEEEeqnarray}{rCl}
\lefteqn{P_{\tilde{\M}_2\tilde{Z}^n}(\hat{\mathcal{A}}_{Z,n}^{\ell_n}) } \quad \nonumber \\
		&= & \sum_{ (x^n,y^n)\in\mathcal{D} } \; P_{\tilde{X}^n\tilde{Y}^n}(x^n,y^n) \nonumber \\	
		 & & \hspace{.8cm}\cdot P_{\tilde{Z}^n|\tilde{X}^n\tilde{Y}^n}(\mathcal{A}_{Z,n}(\phi_2(\phi_1(x^n),y^n)))|x^n,y^n)\IEEEeqnarraynumspace\\
		&\geq&  (1-\zeta_n).\label{eq:zetan}
	\end{IEEEeqnarray}
Defining
	\begin{equation}
		Q_{\tilde{\M}_2}(\m_2) \triangleq \sum_{y^n,\m_1} P_{\tilde{\M}_1}(\m_1)P_{\tilde{Y}^n}(y^n)\cdot \mathbbm{1}\{\phi_2(\m_1,y^n)=\m_2\},
	\end{equation}
we can write
\begin{IEEEeqnarray}{rCl}
	\lefteqn{Q_{\tilde{\M}_2}P_{\tilde{Z}^n}\left(\hat{\mathcal{A}}_{Z,n}^{\ell_n}\right)}\qquad \nonumber\\
		&\leq& Q_{\M_2}P_{Z}^n\left(\hat{\mathcal{A}}_{Z,n}^{\ell_n}\right)\Delta_n^{-3}\\
		& = & \sum_{ \m_2 \in\mathcal{M}_2 } Q_{\M_2} (\m_2)P_{{Z}}^n\left( \hat{\mathcal{A}}_{Z,n}^{\ell_n}(\m_2)\right)\Delta_n^{-3}\IEEEeqnarraynumspace\\
		&\leq& \sum_{ \m_2 \in\mathcal{M}_2 } Q_{\M_2}(\m_2) P_{{Z}}^n\left( {\mathcal{A}}_{Z,n}(\m_2)\right) \nonumber \\
		& & \hspace{1cm} \cdot e^{nh_b(\ell_n/n)}|\mathcal{Z}|^{\ell_n}k_n^{\ell_n}\Delta_{n}^{-3}\IEEEeqnarraynumspace\\
		&=& \beta_{2,n} e^{n \delta_n},\label{Eq:ByCsiszarKornerLemma_lemma}
	\end{IEEEeqnarray}
	where  $\delta_n\triangleq h_b(\ell_n/n) + \frac{\ell_n}{n} \log ( |\mathcal{Z}|\cdot k_n) -\frac{3}{n}\log \Delta_n $ and $k_n \triangleq \min\limits_{\substack{z,z':\\P_Z(z') > 0}}{P_Z(z) \over P_Z(z')}$.
	  Here, (\ref{Eq:ByCsiszarKornerLemma_lemma}) holds by \cite[Proof of Lemma 5.1]{Csiszarbook}.
	
Combining \eqref{Eq:ByCsiszarKornerLemma_lemma} with  \eqref{eq:zetan} and standard inequalities (see \cite[Lemma~1]{JSAIT}), we  then obtain:\vspace{1cm}
		\begin{IEEEeqnarray}{rCl}\label{theta_ub_lemma}
	\lefteqn{-{1\over n}\log \beta_{2,n}} \qquad \nonumber\\
	 & \leq & 	-{1\over n}\log \left( Q_{\tilde{\M}_2}P_{\tilde{Z}^n}\left(\hat{\mathcal{A}}_{Z,n}^{\ell_n}\right) \right) +  \delta_n\\
		&\leq& {1 \over n (1-\zeta_n)} D(P_{\tilde{\M}_2\tilde{Z}^n}||Q_{\tilde{\M}_2}P_{\tilde{Z}^n}) + \delta_n +\frac{1}{n}.\IEEEeqnarraynumspace
	\end{IEEEeqnarray}
	
	We continue to upper bound the divergence term as
	\begin{IEEEeqnarray}{rCl}
		\lefteqn{D(P_{\tilde{\M}_2\tilde{Z}^n}||Q_{\tilde{\M}_2}P_{\tilde{Z}^n})}\qquad \nonumber\\
		&=& I(\tilde{\M}_2;\tilde{Z}^n) + D(P_{\tilde{\M}_2}||Q_{\tilde{\M}_2}) \\
		&\leq& I(\tilde{\M}_2;\tilde{Z}^n) + D(P_{\tilde{Y}^n\tilde{\M}_1}||P_{\tilde{Y}^n}P_{\tilde{\M}_1})\label{eq:dp_ineq_relative_entropy}\\
		&=& I(\tilde{\M}_2;\tilde{Z}^n) + I(\tilde{\M}_1;\tilde{Y}^n)\\
		&=& \sum_{t=1}^n I(\tilde{\M}_2;\tilde{Z}_t|\tilde{Z}^{t-1}) + I(\tilde{\M}_1;\tilde{Y}_t|\tilde{Y}^{t-1})\label{eq:divergence_chainrule}\\
		&\leq& \sum_{t=1}^n I(\tilde{\M}_2\tilde{X}^{t-1}\tilde{Y}^{t-1};\tilde{Z}_t) + I(\tilde{\M}_1\tilde{X}^{t-1}\tilde{Y}^{t-1};\tilde{Y}_t)\IEEEeqnarraynumspace\label{eq:divergence_markovcahins}\\
		&=& \sum_{t=1}^n I(\tilde{V}_{t};\tilde{Z}_t) + I(\tilde{U}_{t};\tilde{Y}_t)\label{eq:divergence_end1}\\
		&=& n[I(\tilde{V}_{T};\tilde{Z}_T|T) + I(\tilde{U}_{T};\tilde{Y}_T|T)]\\
		&\leq& n[I(\tilde{V}_{T}T;\tilde{Z}_T) + I(\tilde{U}_{T}T;\tilde{Y}_T)]\\
		&=& n [I(V;\tilde{Z}) + I(U;\tilde{Y})]\label{theta_ub2_lemma}.
	\end{IEEEeqnarray}
Here \eqref{eq:dp_ineq_relative_entropy} is obtained by the data processing inequality for KL-divergence; \eqref{eq:divergence_chainrule} by the chain rule; \eqref{eq:divergence_markovcahins} by the Markov chain $\tilde{Z}^{t-1} \to (\tilde{X}^{t-1}\tilde{Y}^{t-1}) \to \tilde{Z}_t$; and \eqref{eq:divergence_end1}--\eqref{theta_ub2_lemma} by the definitions of $\tilde{U}_t,\tilde{V}_t,U,V,\tilde{Y},\tilde{Z}$.
	
Following similar steps, we now prove the desired upper bound for $\Pr[\hat{\mathcal{H}}_Y=0|\mathcal{H}=1,(X^n,Y^n)\in\mathcal{D}]$ if \eqref{lem:cond2} is satisfied.
First, note that for any $\eta>0$, \eqref{lem:cond2} implies 
\begin{equation}
\Pr[\hat{\mathcal{H}}_Y=0|\mathcal{H}=0,X^n=x^n,Y^n=y^n]= 1, \;\; \forall (x^n,y^n) \in \mathcal{D},
\end{equation} 
due to the fact that given a pair of sequences $(x^n,y^n)$, the probability that {\Rel} decides $\hat{\mathcal{H}}_Y=0$ is either 0 or 1. Define the acceptance region  at {\Rel} as
\begin{equation}
 \mathcal{A}_{Y,n} \triangleq \{ (\m_1,y^n)\colon g_1(\m_1,y^n) =0 \}.
\end{equation}
Then,
\begin{IEEEeqnarray}{rCl}
	P_{\tilde{\M}_1\tilde{Y}^n}({\mathcal{A}}_{Y,n})
	&=&  \sum_{\substack{(x^n,y^n)\in\mathcal{D},\\\m_1=\phi_1(x^n),\\(\m_1,y^n)\in{\mathcal{A}}_{Y,n}}} P_{\tilde{X}^n\tilde{Y}^n\tilde{\M}_1}(x^n,y^n,\m_1)\IEEEeqnarraynumspace\\[1.2ex]
	&=&  1,\label{eq:12}
\end{IEEEeqnarray}
and inspecting the definitions of $P_{\tilde{M}_1\tilde{Y}^n}$ and $\beta_{1,n}$:
\begin{IEEEeqnarray}{rCl}
	P_{\tilde{\M}_1}P_{\tilde{Y}^n}\left({\mathcal{A}}_{Y,n}\right)
		&\leq& P_{{\M}_1}P_{{Y}^n}\left({\mathcal{A}}_{Y,n}\right)\Delta_n^{-2}\\
				&=& \beta_{1,n} \Delta_n^{-2}. \label{eq:11}
	\end{IEEEeqnarray}
By \eqref{eq:11}, \eqref{eq:12} and standard inequalities (see \cite[Lemma~1]{JSAIT}), we  further obtain 
\begin{IEEEeqnarray}{rCl}\label{theta_ub_lemma_relay}
-{1\over n}\log \beta_{1,n} & \leq &-{1\over n}\log \left( P_{\tilde{\M}_1}P_{\tilde{Y}^n}\left({\mathcal{A}}_{Y,n}\right)  \right) - \frac{2}{n} \log \Delta_n \IEEEeqnarraynumspace\\
		&\leq& {1 \over n } D(P_{\tilde{\M}_1\tilde{Y}^n}||P_{\tilde{\M}_1}P_{\tilde{Y}^n}) + \delta_n'\IEEEeqnarraynumspace
\end{IEEEeqnarray}
where $\delta_n'\triangleq - \frac{2}{n} \log \Delta_n +\frac{1}{n}$ and tends to 0 as $n \to \infty$.

We continue to upper bound the divergence term as
\begin{IEEEeqnarray}{rCl}
	D(P_{\tilde{\M}_1\tilde{Y}^n}||P_{\tilde{\M}_2}P_{\tilde{Y}^n})	&=& I(\tilde{\M}_1;\tilde{Y}^n)\\
	&=& \sum_{t=1}^n I(\tilde{\M}_1;\tilde{Y}_t|\tilde{Y}^{t-1})\label{eq:divergence_chainrule_relay}\\
	&\leq& \sum_{t=1}^n I(\tilde{\M}_1\tilde{X}^{t-1}\tilde{Y}^{t-1};\tilde{Y}_t)\IEEEeqnarraynumspace\label{eq:divergence_markovcahins_relay}\\
	&=& \sum_{t=1}^n I(\tilde{U}_{t};\tilde{Y}_t)\label{eq:divergence_end1_relay}\\
	&=& n[I(\tilde{U}_{T};\tilde{Y}_T|T)]\\
	&\leq& n[I(\tilde{U}_{T}T;\tilde{Y}_T)]\\
	&=& n [I(U;\tilde{Y})]\label{theta_ub2_lemma_relay}.
\end{IEEEeqnarray}
Here \eqref{eq:divergence_chainrule_relay} holds by the chain rule and \eqref{eq:divergence_end1_relay}--\eqref{theta_ub2_lemma_relay} hold by the definitions of $\tilde{U}_t,U,\tilde{Y}$.

Finally, we proceed to prove the Markov chain $U \to \tilde{X} \to \tilde{Y}$ in the limit as $n \to \infty$. To this end, notice the Markov chain $\tilde{\M}_1 \to \tilde{X}^n \to \tilde{Y}^n$, and thus similar to the analysis in \cite[Section V.C]{HWS20}:
	\begin{IEEEeqnarray}{rCl}
		0 &=& I(\tilde{\M}_1;\tilde{Y}^n|\tilde{X}^n) \\ 
		&\geq& H(\tilde{Y}^n|\tilde{X}^n)  - H(\tilde{Y}^n|\tilde{X}^n\tilde{\M}_1) \nonumber \\
		&& + D(P_{\tilde{X}^n\tilde{Y}^n}||P_{XY}^n) + \log{\Delta_{n}}
		\label{MC1proofstep1}\\
		&{\geq}& n[H(\tilde{Y}_T|\tilde{X}_{T}) + D(P_{\tilde{X}_{T}\tilde{Y}_T}||P_{XY})] +\log{\Delta_{n}} \nonumber\\
		&& - H(\tilde{Y}^n|\tilde{X}^n\tilde{\M}_1) \label{MC1proofstep2}\\
		&\geq& n[H(\tilde{Y}_T|\tilde{X}_{T}) + D(P_{\tilde{X}_{T}\tilde{Y}_T}||P_{XY})] +\log{\Delta_{n}} \nonumber \\ 	&&-  \sum_{t=1}^{n}H(\tilde{Y}_t|\tilde{X}_{t}\tilde{X}^{t-1}\tilde{Y}^{t-1}\tilde{\M}_1)\label{MC1proofstep3}\\
		&=& n[H(\tilde{Y}_T|\tilde{X}_{T}) + D(P_{\tilde{X}_{T}\tilde{Y}_T}||P_{XY})] +\log{\Delta_{n}} \nonumber \\
		&&-  \sum_{t=1}^{n}H(\tilde{Y}_t|\tilde{X}_{t}\tilde{U}_{t})\label{MC1proofstep4}\\
		&\geq& n[H(\tilde{Y}_T|\tilde{X}_{T}) - H(\tilde{Y}_T|\tilde{X}_T,\tilde{U}_{T},T) ]+ \log{\Delta_{n}}  \IEEEeqnarraynumspace\label{MC1proofstep4b}\\
		&\geq& nI(\tilde{Y};{U}|\tilde{X}) + \log{\Delta_{n}},\label{MC1proofstep5}
	\end{IEEEeqnarray}
	where \eqref{MC1proofstep2} holds by the super-additivity property in \cite[Proposition 1]{tyagi2019strong}; \eqref{MC1proofstep3} by the chain rule and since conditioning reduces entropy; \eqref{MC1proofstep4} by the definition of $\tilde{U}_{t}$; \eqref{MC1proofstep4b} by the non-negativity of the Kullback-Leibler divergence, and by recalling that $T$ is uniform over $\{1,\ldots,n\}$ independent of all other random quantities, and finally \eqref{MC1proofstep5} holds by the definitions of ${U},\tilde{X}, \tilde{Y}$.

\bibliographystyle{ieeetr}
\bibliography{references}
\end{document}